# Looping metal-support interaction in heterogeneous catalysts during redox reactions

Yue Pan[1,2†], Shiyu Zhen[3†], Xiaozhi Liu[1†], Mengshu Ge[1], Jianxiong Zhao[1], Lin Gu[4],

Dan Zhou[5*], Liang Zhang[3*], Dong Su[1,2*]

[1]*Beijing National Laboratory for Condensed Matter Physics, Institute of Physics, Chinese Academy of Sciences, Beijing, China.*
[2]*University of Chinese Academy of Sciences, Beijing, China.*
[3]*Center for Combustion Energy, School of Vehicle and Mobility, Tsinghua University, Beijing, China*
[4]*Beijing National Center for Electron Microscopy and Laboratory of Advanced Materials, School of Materials Science and Engineering, Tsinghua University, Beijing, China.*
[5]*Leibniz Institute for Crystal Growth, Berlin, Germany*

[†]These authors contributed equally to this work.
[*]Corresponding authors. E-mail: *dan.zhou@ikz-berlin.de (D.Z.)*, *zhangbright@tsinghua.edu.cn (L.Z.)*, *dongsu@iphy.ac.cn (D.S.)*

**Abstract**

Metal-support interfaces fundamentally govern the catalytic performance of heterogeneous systems through complex interactions. Here, utilizing operando transmission electron microscopy, we uncovered a type of looping metal-support interaction in NiFe-$Fe_3O_4$ catalysts during hydrogen oxidation reaction. At the NiFe-$Fe_3O_4$ interfaces, lattice oxygens react with NiFe-activated H atoms, gradually sacrificing themselves and resulting in dynamically migrating interfaces. Meanwhile, reduced iron atoms migrate to the {111} surface of $Fe_3O_4$ support and react with oxygen molecules. Consequently, the hydrogen oxidation reaction separates spatially on a single nanoparticle and is intrinsically coupled with the redox reaction of the $Fe_3O_4$ support through the dynamic migration of metal-support interfaces. Our work provides previously unidentified mechanistic insight into metal-support interactions and underscores the transformative potential of operando methodologies for studying atomic-scale dynamics.

**Main**

Heterogeneous catalysts play a pivotal role in a variety of chemical processes, spanning chemical synthesis, energy conversion, and environmental remediation[1-5]. In metal-oxide support systems, oxide serves multiple critical functions: stabilizing metal particles under extreme conditions and modulating catalytic behavior through metal-support interactions (MSIs)[6-9]. These interactions primarily governed by the metal-metal or metal-oxygen interactions at the interface[10], can result in complex phenomena such as electronic metal-support interaction (EMSI), strong metal-support interaction (SMSI), and reactive metal-support interaction (RMSI)[11-14]. Such MSIs profoundly influence catalytic performance, directly impacting critical parameters like reaction activity, selectivity, and stability across diverse reactions, including hydrocarbon dehydrogenation, Fischer-Tropsch synthesis, catalytic oxidation, gas/liquid phase reforming, and selective hydrogenation[15-18]. Therefore, understanding interfacial structures and gaining precise control over MSIs has emerged as an important perspective for optimizing catalytic performances[19,20].

Recent advancements in environmental transmission electron microscopy (ETEM) technologies enable real-time observation of catalyst structural evolutions during reactions, providing atomic-scale insights into reaction mechanisms[18,21,22]. These studies have revealed the highly dynamic nature of catalyst structures, which are significantly influenced by the surrounding chemical environment[11,23-26]. In-situ ETEM investigations of Pt-$TiO_2$ systems have revealed remarkable gas-induced oscillations at the metal-support interface, including dramatic structural collapses and reconstructions under redox conditions[21,27]. These transformations are believed to originate from the cyclic generation and refilling of oxygen vacancies. However, despite these results, the fundamental dynamics of MSIs and how their interfacial structure governs catalytic reactions remain poorly understood across many important catalytic systems.

In this work, we investigate the reaction dynamics of the NiFe-$Fe_3O_4$ catalyst during hydrogen oxidation ($H_2 + O_2 \rightarrow H_2O$)[24,27] using the ETEM technique and uncover a novel looping metal-support interaction (LMSI). Yielding only water as the product in gas phase reactions, the hydrogen oxidation reaction serves as a critical model system for investigating fundamental processes such as adsorption and surface reactions on heterogeneous catalysts[28-30]. Moreover, the coexistence of hydrogen and oxygen simulates redox environments relevant to various catalytic reactions, including $CO_2$ hydrogenation, CO oxidation, methane combustion, and the water-gas shift reaction, particularly for transition metal catalysts like Ni and Fe, which are readily oxidized and reduced under redox conditions[31-33]. To construct the active interface, we reduce $NiFe_2O_4$ (NFO) with hydrogen, thereby forming a $NiFe$-$Fe_3O_4$ structure that enables us to investigate the metal-support interaction under redox conditions.

By controlling the redox environment through gas composition and temperature, we directly visualize reaction-induced solid-solid interfacial migration under reaction conditions and reveal the following mechanism: hydrogen spillover on NiFe nanoparticles provide $H^+$ to the $NiFe$-$Fe_3O_4$ interface, triggering the reduction of $Fe_3O_4$. Reduced $Fe^0$ adatoms then migrate substantial distances to the edge sites of $Fe_3O_4$ {111} planes, where they facilitate oxygen molecule activation. In this process, the redox of iron is coupled with hydrogen oxidation under the catalytic effect of the NiFe nanoparticle. Through quantitative analyses and theoretical calculations, we establish a direct correlation between metal-support interfacial dynamics and catalytic activity. This work highlights the role of LMSI in enhancing catalytic performance and provides novel insights into the fundamental processes of heterogeneous catalysis.

**Result**

As illustrated in **Supplementary Fig. 1**, we investigate the MSI of $NiFe$-$Fe_3O_4$ catalysts using a gas-celled ETEM equipped with a quadrupole mass spectrometer (MS). The $NiFe$-$Fe_3O_4$ catalyst is synthesized through the partial reduction of the NFO precursor (**Supplementary Fig. 2 and 3)**. Following a reduction procedure in 10% $H_2$/He at 400 °C, the NFO nanoparticles transform into a composition of NiFe and $Fe_3O_4$, as confirmed by select area electron diffraction (SAED) analysis in **Supplementary Fig. 4**. The structural characterization of the synthesized $NiFe$-$Fe_3O_4$, presented in **Supplementary Fig. 5**, reveals a classical SMSI state, with NiFe nanoparticles encapsulated by $FeO_x$ layers. Notably, these encapsulated layers exhibit remarkable instability in redox environments, completely retracting from the NiFe nanoparticles upon introduction of the reactant gas mixture (2% $O_2$, 20% $H_2$, and 78% He) into the gas cell (**Supplementary Fig. 6**). This gas-induced destabilization of the encapsulating layers is analogous to behaviors observed in other metal-reducible oxide support system, where oxygen abstraction from the overlayer under redox conditions leads to dynamic metal-support interactions[27,34-36].

When the reaction temperature exceeds 500 °C, a distinctive catalytic dynamic behavior emerges, characterized by the coordinated migration of NiFe nanoparticles across the $Fe_3O_4$ support, accompanied by the etching and reconstruction of the support material (**Supplementary Video 1**). In contrast, in-situ experiments on pure Ni and $Fe_3O_4$ catalysts during hydrogen oxidation reaction reveal no comparable dynamic behavior (**Supplementary Fig. 7**). This dynamic behavior is consistently observed across all NiFe nanoparticles in the NiFe-$Fe_3O_4$ system, suggesting a significant synergetic interaction between the NiFe nanoparticles and the $Fe_3O_4$ support, which dynamically responds to the redox reaction environment. We propose the term looping metal-support interaction (LMSI) to describe this phenomenon. The LMSI is primarily driven by high-pressure, high-temperature conditions typical of work conditions, which simulate the chemical potential of $H_2$ and $O_2$ required to promote the structural evolution of the catalysts. These reaction conditions, and the associated structural dynamics they provoke, highlight the critical importance of studying catalyst behavior under authentic operational conditions through real-time operando TEM observations.

HRTEM sequence images captured at 700 °C depict the interface reactions associated with LMSI, as shown in **Supplementary Video 2** and **Fig. 1a** to **f**. The images reveal a preferential epitaxial relationship between NiFe nanoparticles and $Fe_3O_4$ support during the dynamic process. Based on the FFT analysis in **Fig. 1g** and the simulated SAED pattern in **Supplementary Fig. 8**, the orientational relationship between the NiFe and $Fe_3O_4$ particles is determined to be: NiFe (1-12) // $Fe_3O_4$ (1-1-1) and NiFe [110] // $Fe_3O_4$ [110]. This correlation is further validated by simulating the HRTEM image of the interface structure, which aligns closely with experimental results (**Fig. 1h, Supplementary Figs. 9** and **10**). Interestingly, the lattice spacing of NiFe (-111) is 0.20 nm, approximately 15% larger than that of $Fe_3O_4$ (2-24) (0.17 nm). This lattice mismatch results in a 4.2° tilting of the NiFe (1-12) plane relative to the $Fe_3O_4$ (1-1-1) plane, a configuration that minimizes interfacial strain[37-39]. Consequently, this tilting leads to the formation of lattice voids along the interface, as illustrated in the structural model presented in **Supplementary Fig. 11**.

The epitaxial NiFe-$Fe_3O_4$ interface initially forms a ledge-and-terrace configuration as denoted by the black dashed line in **Fig. 1a**. Subsequently, the interface migrates through lateral propagation of atomic ledge along the NiF-$Fe_3O_4$ interface (**Fig. 1b to f**). The interface migration process led to the layer-by-layer dissolution of the $Fe_3O_4$ along the (111) plane, preserving the continuity of the interface structure. The dissolution resembles the Mars-van Krevelen mechanism, where activated hydrogen from NiFe spills over to the interface, facilitating the release of lattice oxygen and reduction of $Fe_3O_4$[40,41]. During this process, surface Ni (Fe) atoms transit to lattice voids placed at the interface, driving the migration of NiFe toward the $Fe_3O_4$ side (**Supplementary Fig. 12**). Therefore, the vacancies generated by the exsolution of Fe and O from the $Fe_3O_4$ support are promptly replenished by migrating NiFe atoms, resulting in a self-regulated dynamic interface that minimizes the accumulation of persistent defects. As surface atoms diffuse, the morphology of NiFe deforms with the propagation of NiFe-$Fe_3O_4$ interface, as illustrated in **Supplementary Fig. 13**. Although Wulff construction predicts that NiFe nanoparticles would expose {111}, {100}, {311}, {110} surfaces[42], surface-diffusion-induced dynamics result in varying proportions of surface terminations. The metal nanoparticle morphology, induced by interface dynamics, enables NiFe to continuously expose high-index active sites for hydrogen activation. Notably, as the NiFe particle adjusts to achieve lattice matching, its shape changes like a liquid droplet while maintaining its single-crystalline structure[43].

With lattice oxygen released from $Fe_3O_4$ at the interface, $Fe^{2/3+}$ ions are reduced to mobilizable adatoms that can migrate across the $Fe_3O_4$ surface, capturing and activating oxygen gas molecules. The activation of oxygen gas takes place at the $Fe_3O_4$ {111} facets, away from the NiFe-$Fe_3O_4$ interface, as illustrated in **Supplementary Video 3**. As shown in **Fig. 1i**, the $Fe_3O_4$ predominantly exposes the atomically flat {111} facets. Upon reaction between reduced Fe adatoms and oxygen gas molecules, a monoatomic layer nucleates at the flat {111} plane of the $Fe_3O_4$, creating a step-terrace configuration, as marked by the red arrows in **Fig. 1j**. This newly formed atomic layer then grows laterally through step-flow growth with Fe and O adatoms attaching to the step edge (**Fig. 1k to n**). The Fe atoms, originating from the NiFe-$Fe_3O_4$ interface and diffusing outward on the $Fe_3O_4$ support, facilitate the dissociative adsorption of $O_2$ molecules[23]. The growth behavior ensures that the $Fe_3O_4$ support continuously exposes {111} planes, lowering the system's energy. Overall, the LMSI of NiFe-$Fe_3O_4$ catalysts align with gas-solid reactions at separated sites, including the reduction of $Fe_3O_4$ at NiFe-$Fe_3O_4$ interface and oxidation of Fe adatoms at $Fe_3O_4$ {111} sites, as illustrated in **Fig. 1o**. These two gas-solid reactions are synergistically facilitated by the migration of Fe adatoms.

TEM images **(Supplementary Fig. 14)** and the video **(Supplementary Video 1)** recorded at 700 °C reveal that the LMSI between NiFe and $Fe_3O_4$ support significantly contributes to its structural dynamics. The simultaneous reduction and oxidation of $Fe_3O_4$ are governed by the chemical potential of the gas phase, which drives the system into a nonequilibrium dynamic state[23,24,44,45]. As shown in **Supplementary Fig. 15**, we investigate the NiFe-$Fe_3O_4$ catalysts under varying $H_2/O_2$ ratios and find that increasing the oxygen content suppresses redox dynamics, necessitating higher temperatures to initiate interface reduction and surface oxidation processes.

Time-sequence TEM images of a NiFe-$Fe_3O_4$ system in **Fig. 2a to h** and **Supplementary Video 4** illustrate the synergistic between interfacial reduction and surface oxidation of $Fe_3O_4$. The phases of NiFe and $Fe_3O_4$ are confirmed by the corresponding filtered inverse Fast Fourier Transform (iFFT) presented in **Fig. 2i**. To clearly illustrate interface migration and surface growth, the NiFe nanoparticle and growth $Fe_3O_4$ layers are denoted by blue and red colors, respectively. It is evident that as the interface migrates, the $Fe_3O_4$ support is etched, exposing a pathway, while simultaneously, $Fe_3O_4$ layers grow at the {111} planes near the interface. In **Fig. 2j**, we track the trajectory of the NiFe nanoparticle, revealing its movement along the <112> direction of $Fe_3O_4$. This directional movement correlates with layer-by-layer etching of the $Fe_3O_4$ at {111} planes. Additionally, the etching process exposes new {111} planes of $Fe_3O_4$ along the path of the NiFe nanoparticle. Upon reaching the boundary of the $Fe_3O_4$, the NiFe nanoparticle reorients and continues its movement in the specific direction (**Fig. 2e** to **f**, **Supplementary Fig. 13** and **16**). The directional migration of NiFe nanoparticles, along with the specific lattice growth of $Fe_3O_4$, ensures that the $Fe_3O_4$ consistently exposes low index {111} planes, thereby minimizing the entire energy of the catalyst system[46,47]. We have recorded the movement of different nanoparticles (**Supplementary Fig. 17**) confirming the above-mentioned conclusions. We find that the reduction and oxidation processes are coupled, forming an integral part of the overall hydrogen oxidation reaction. Although reduction and oxidation occur at separate sites, the proportions of $Fe_3O_4$ and NiFe remain steady, as evidenced by the measured projection area in **Fig. 2k**, further confirming the coupled iron redox process.

To further clarify the correlation between the LMSI and catalytic activity, we simultaneously measured the MS signals of oxygen and water in **Fig. 3a**. The results reveal the production of $H_2O$ and the consumption of $O_2$ at temperatures above 500 °C, suggesting that the hydrogen oxidation reaction occurs. In contrast, blank experiments using empty chips (**Supplementary Fig. 18**) show no variation in signal with temperature increases up to 900 °C, confirming that the detected water signal originates from catalytic H2-O2 reaction. Additionally, sole Fe3O4 nanoparticles exhibit no catalytic activity nor LMSI effect under the same conditions (**Supplementary Fig. 7 and 19**). In-situ SAED pattern recorded during the structural dynamics (**Fig. 3b**) and corresponding peaks analysis (**Fig. 3c**) suggest that the observed structural dynamics arise from phase transitions between NiFe and Fe3O4. We simultaneously record the structural evolution of the NiFe-Fe3O4 catalyst during the above reaction process, as illustrated in **Fig. 3d** to **h** and **Supplementary Videos 5-9**. At the onset of the reaction at 500 °C, LMSI effects in the NiFe-Fe3O4 nanoparticles are observed (**Supplementary Figs. 20** and **21**). As the temperature increases to 600 °C, water production rises significantly, accompanied by more pronounced LMSI effects of the nanoparticles (**Supplementary Video 7**). The enhanced structural dynamics are also accompanied by an increase in NiFe particle size by both Ostwald ripening and coalescence process (**Supplementary Fig. 22**). However, the growth becomes limited as the temperature exceeds 700 °C, suggesting that particle sintering is constrained by dynamic interface interactions.

To quantify these dynamics, we have employed an optical flow method to quantitatively track the movement velocity of nanoparticles[48]. As shown in **Fig. 3i** to **m**, the velocity of particle movement increases notably from 400 °C to 600 °C, reflecting a consistent trend with catalytic activity. With the reaction temperature reaching 800 °C, both MS signals and the velocity of particle movement reach steady. In-situ SAED patterns are also recorded from 400 °C to 800 °C (heat rate: 1 °C/s) to illustrate the structural dynamics above 500 °C (**Supplementary Video 10**). A diffraction-spot-detection method[49] is employed to track these dynamic diffraction spots (**Supplementary Video 11**), revealing a noticeable reduction in their lifespan (**Fig. 3n**). Our statistics analysis of the average velocity magnitudes of the structural dynamics and the average lifespan of the dynamic diffraction spots (**Fig. 3o**) further corroborates the close relationship between LMSI and catalytic activity.

We investigate the LMSI effect on the spatially decoupled hydrogen oxidation reaction in the NiFe-Fe3O4 system through DFT calculations with Hubbard U correction, as shown in **Fig. 4a**. The computed Gibbs free energy diagram reveals an exceptionally low hydrogen dissociation barrier of 0.11 eV for the NiFe-Fe3O4 system, which is significantly lower than the barriers observed for Fe3O4 and Fe-Fe3O4 systems (**Supplementary Fig. 23**). The activated hydrogen then spillover to the interface, promoting the reduction of the Fe3O4 support, as discussed in detail by Prins, R[50]. NN-MD simulations further elucidate the interface reduction mechanism, demonstrating that NiFe preferentially attracts oxygen from the $Fe_3O_4$ due to its stronger oxygen affinity compared to Fe-$Fe_3O_4$ (**Fig. 4b, 4c** and **Supplementary Figs. 24-26**). This phenomenon, termed the oxygen reverse spillover effect, facilitates the reduction of $Fe_3O_4$ layers at the interface[51] (**Supplementary Text 3**). This observed tendency of oxygen attraction to NiFe aligns closely with our experimental findings from reduction and oxidation experiments on the NFO sample (**Supplementary Text 4 and 5**). The attracted oxygen atoms on the NiFe surface are subsequently reduced by dissociated hydrogen atoms, forming water through a process that is substantially more efficient than on pure Fe, where hydrogen activation and water formation typically constitute rate-limiting steps (**Supplementary Figs. 27-29**).

Concurrently, the reduced Fe atoms migrate to the $Fe_3O_4$ {111} facets, where they undergo re-oxidization by oxygen. This migration proceeds significantly more rapidly on the surface than through the $Fe_3O_4$ bulk (**Fig. 4d**). Besides, the NiFe's superior $H_2$ dissociation activity enables hydrogen spillover across a broader region on the NiFe-$Fe_3O_4$ system, creating an extended hydrogen-rich environment. As a result, Fe atoms in the NiFe-$Fe_3O_4$ system need to traverse a longer distance before re-oxidation compared to those in the Fe-$Fe_3O_4$ system. This extended Fe diffusion pathway is fundamental to the spatially decoupled redox process characteristic of NiFe-$Fe_3O_4$, which distinguishes it from the more localized redox behavior seen in Fe-$Fe_3O_4$. The energy profile illustrating the dissociation pathway of molecular oxygen ($O_2$) on an Fe adatom anchored to the $Fe_3O_4$ surface (**Supplementary Fig. 30**) shows that the energy barrier is low, indicating that this step is not rate-limiting in the redox cycle. **Fig. 4e** provides a comprehensive visualization of the correlation between LMSI and the hydrogen oxidation mechanism in NiFe-$Fe_3O_4$. The dynamic process is characterized by spatially decoupled hydrogen and oxygen activation (hydrogen activation occurs around NiFe, while oxygen activation takes place at distant $Fe_3O_4$ {111} facets). This is accompanied by $Fe_3O_4$ reduction, Fe atom migration, and subsequent re-oxidation. The low $H_2$ dissociation barrier on NiFe, combined with rapid surface diffusion of Fe atoms, facilitates efficient hydrogen spillover and swift Fe atom migration. These factors are crucial in enabling spatial separation that define the unique LMSI phenomenon in the NiFe-$Fe_3O_4$ system.

In summary, our comprehensive investigation using operando ETEM reveals a new type of LMSI in the NiFe-$Fe_3O_4$ catalyst system. The $Fe^{2/3+}/Fe^0$ redox cycles demonstrate remarkable spatial segregation, activating hydrogen and oxygen at distinctly different locations. This mechanism drives two critical processes: reduction (etching) of $Fe_3O_4$ at the NiFe-$Fe_3O_4$ interface and promoting growth of $Fe_3O_4$ layers along the {111} planes. The LMSI effect emerges from the inherent reducibility of the $Fe_3O_4$ support, with interface migration precisely orchestrated by the movement of reduced iron species. NiFe nanoparticles play a role in prompting the spillover of hydrogen and accelerating the iron redox process. By elucidating the intricate interplay between catalyst components, we establish a fundamental correlation between the LMSI effect and catalytic performance. Beyond this specific case, we suggest a broader applicability of LMSI to heterogeneous catalytic systems containing transition metal-oxide support. The hydrogen spillover can occur on metallic nanoparticles, coupling with the redox circles of other types of transition metal-oxide support, consequently enhancing overall catalytic reactivity. By unveiling the sophisticated spatial and chemical dynamics at the catalyst interface, our research provides a transformative perspective on metal-support interactions, potentially guiding the design of highly efficient heterogeneous catalysts.

# Figures

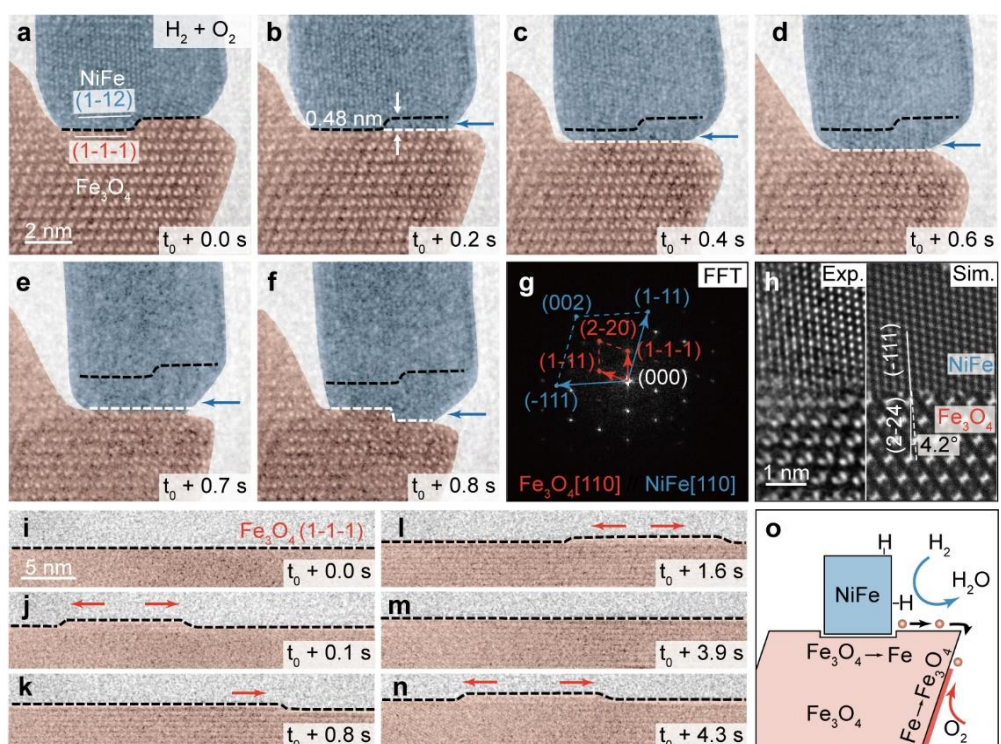

**Fig. 1. Atomic-resolution observation of the LMSI of NiFe-Fe$_3$O$_4$ catalyst. a-f**, Time-sequence HRTEM images of a NiFe-Fe$_3$O$_4$ nanoparticle under hydrogen oxidation reaction condition at 700 °C from **Supplementary Video 2**, showing surface-diffusion-induced migration of the NiFe nanoparticle, along with the layer-by-layer reduction of Fe$_3$O$_4$. **g**, FFT pattern of **a**, illustrating the epitaxial relationship between the NiFe and Fe$_3$O$_4$, i.e., NiFe (1-12) // Fe$_3$O$_4$ (1-1-1) and NiFe [110] // Fe$_3$O$_4$ [110]. **h** is an enlarged image of a selected region in **a**. In comparison, a simulated HRTEM image is also shown. **i-n**, Observed layer-by-layer growth of Fe$_3$O$_4$ at the surface of Fe$_3$O$_4$ from **Supplementary Video 3**. **o**, Schematic image illustrates the relation between the LMSI effect and catalytic reactions. The reduction of Fe$_3$O$_4$ at the NiFe-Fe$_3$O$_4$ interface resulted in the migration of the interface. Additionally, the oxidation of Fe at the edge sites of the Fe$_3$O$_4$ {111} planes leads to the growth of Fe$_3$O$_4$.

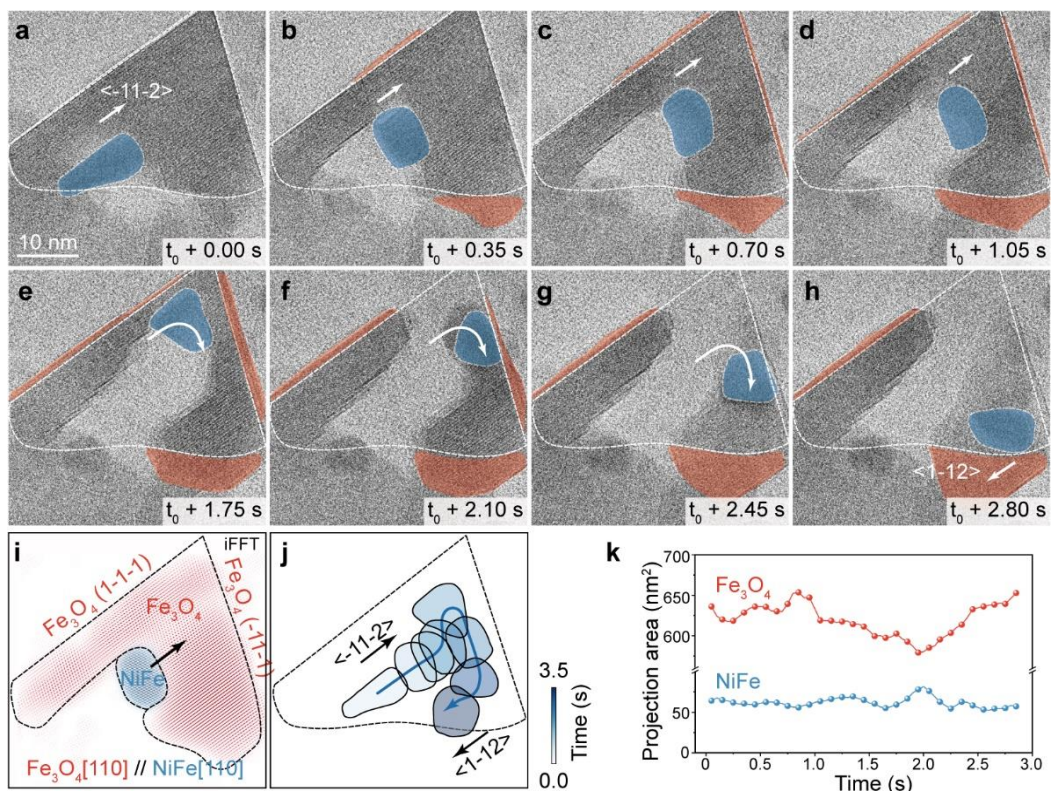

**Fig. 2. Dynamic equilibrium of LMSI in NiFe-Fe$_3$O$_4$ catalysts during hydrogen oxidation reaction. a-h**, Time-sequence HRTEM images of NiFe-Fe$_3$O$_4$ catalyst at 700 °C at an O$_2$/H$_2$ ratio of 1/10 in **Supplementary video 4**. **i**, Filtered inverse FFT of HRTEM image in **b** shows the correlated orientation between Fe$_3$O$_4$ and NiFe. **j**, Trace of motion path and morphology evolution of the NiFe nanoparticle in **Supplementary video 4**. **k**, Projection area of Fe$_3$O$_4$ and NiFe in function of time from **Supplementary video 4**.

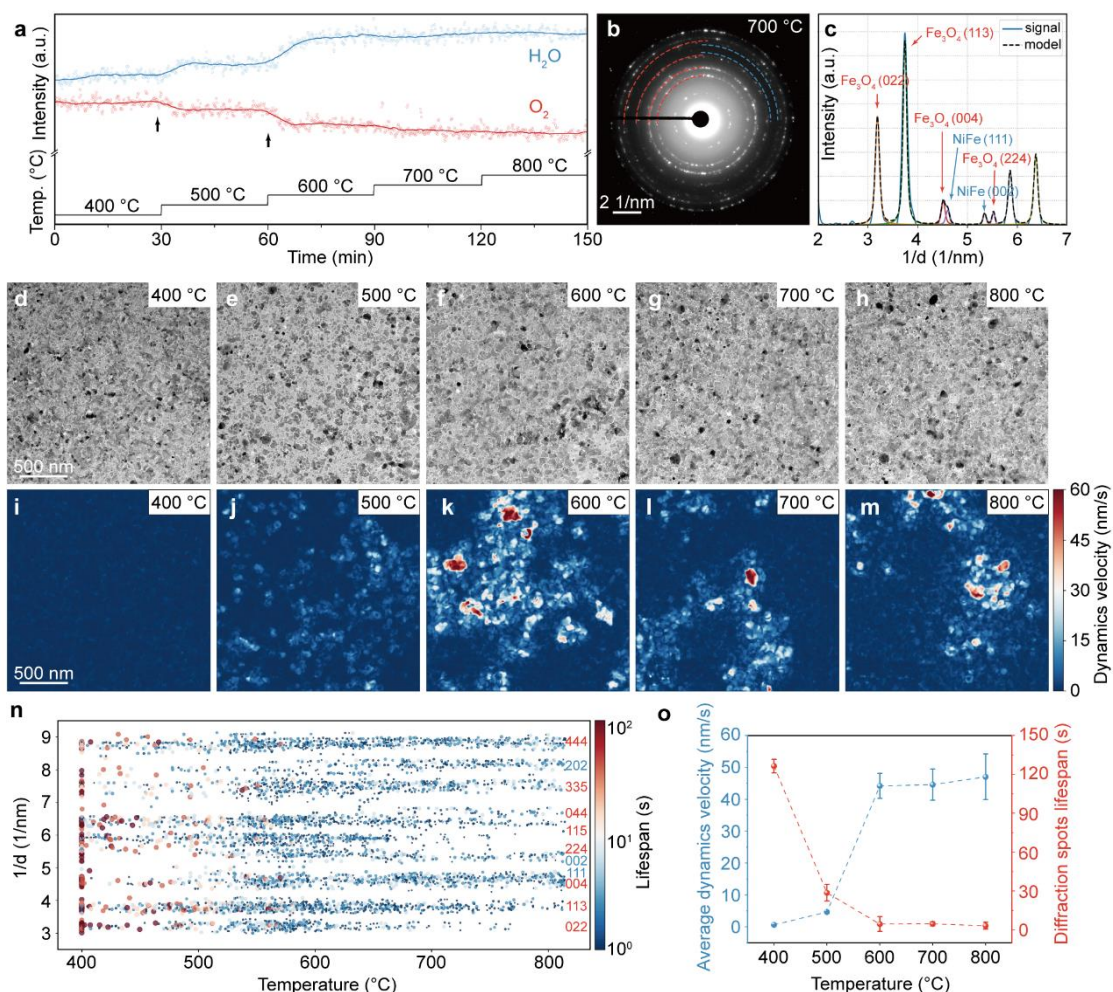

**Fig. 3. Correlation between the LMSI and catalytic activity of NiFe-Fe$_3$O$_4$ catalyst in hydrogen oxidation reaction. a**, Operando MS data obtained post-TEM at temperatures from 400°C to 800°C. **b-c**, SAED pattern and corresponding peaks analysis of the catalyst. **d-h**, Representative TEM images of NiFe-Fe$_3$O$_4$ nanoparticles at different temperatures captured from **Supplementary Video 5-9**. **i-m**, Dynamics velocity analyses with an optical flow method based on **Supplementary Video 5-9**. **n**, Lifespan from tracking diffraction spots of in-situ SAED (**Supplementary Video 10** and **11**) pattern captured with heating under ambient conditions. The reciprocal spacing and lifespan of each dot are shown in the function of temperature. **o**, The average dynamics velocity (blue dots), and the lifespan of diffraction spots (red dots) across various temperature stages illustrate a close correlation between structural dynamics and catalytic activity.

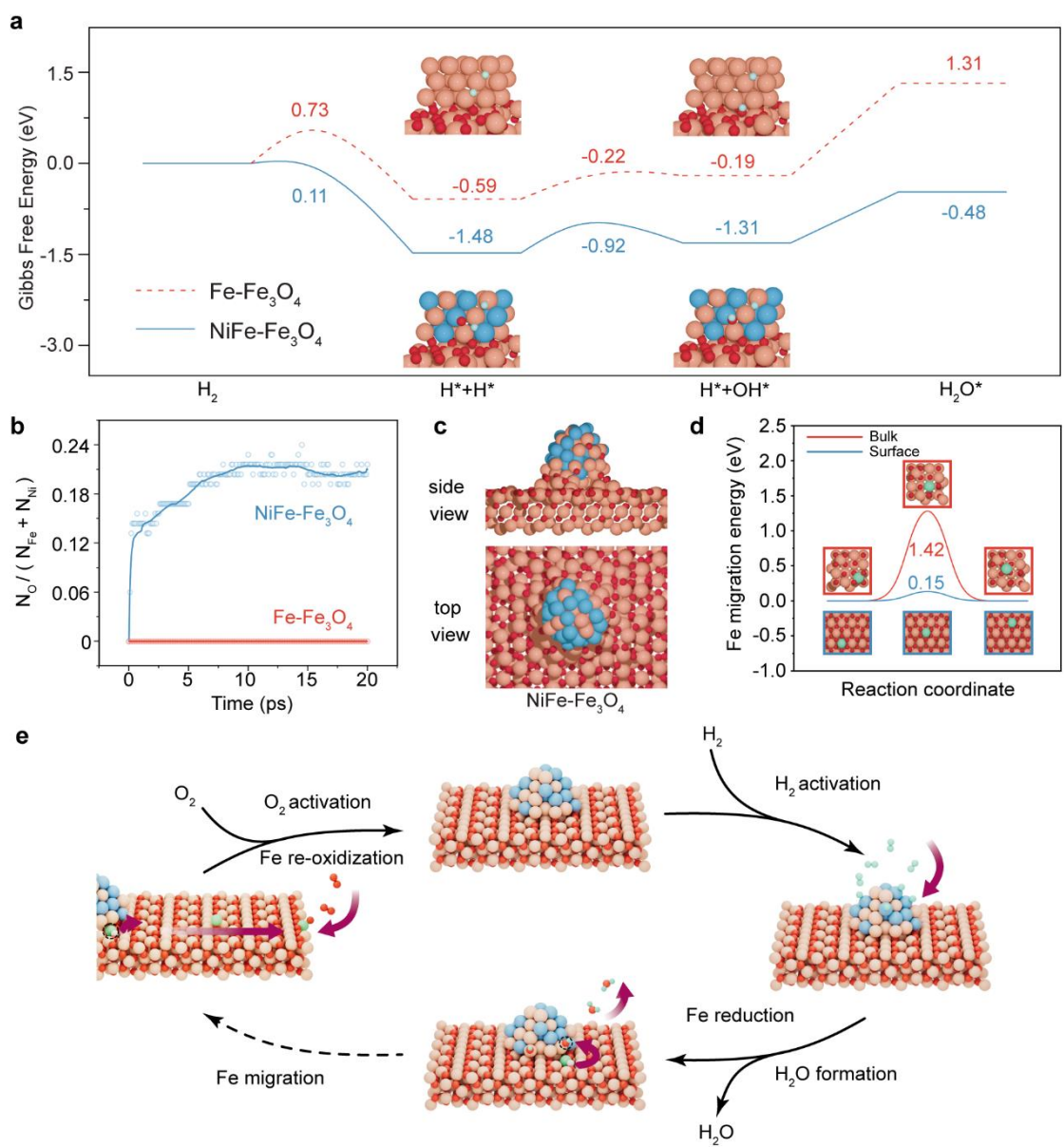

**Fig. 4. Theoretical analysis of the atomic reaction pathway on the NiFe-Fe₃O₄ catalyst. a**, DFT calculated Gibbs free energy diagram for the hydrogen oxidation reaction on NiFe-Fe₃O₄ (blue solid), and Fe-Fe₃O₄ (red dashed). Insets show optimized structures of intermediate states. Cyan, pink, and red, water balls display Ni, Fe, O, and H atoms, respectively. **b**, Amount of accumulated oxygen atoms on NiFe and Fe nanoparticles from NN-MD simulations, showing that NiFe facilitates the migration of lattice oxygen from the Fe₃O₄ to the NiFe nanoparticle. **c**, Side and top view of NiFe-Fe₃O₄ after 20 ps NN-MD simulation. **d**, Energy profile for activated Fe atom transfer on Fe₃O₄. **e**, Schematics representation of the mechanisms underlying the LMSI effect in hydrogen oxidation on the NiFe-Fe₃O₄ catalyst.

**Methods**

Operando TEM setup

The operando TEM experiments were conducted using a JEM-F200, a DENSsolutions Climate G+ gas supply system, and a quadrupole mass spectrometer (**Supplementary Fig. 1**). It has been shown that the purity of the gas composition significantly affects the state of nanoparticles, and optimizing the environment in ETEM can closely simulate actual working conditions. To prevent contamination of pipelines, we construct our gas supply system using stainless steel or poly(ether-ether-ketone)–silica, minimizing leakage and moisture adsorption. Gas filters tailored for different gas sources (including hydrogen, oxygen, and helium) are also implemented to remove various impurities, such as oxygen, moisture, and hydrocarbon contaminants. Additionally, a complete workflow for operando experiments is proposed, including leakage testing and pipeline flushing, to ensure precise control of the microenvironment in the gas-cell nanoreactor.

After completing the preparation procedures, NFO samples are dispersed in ethanol with sonication, then drop-cast onto a microelectromechanical system (MEMS)-based heating chip. This heating chip is sealed with another chip featuring a large electron-transparent $Si_3N_4$ membrane for imaging, forming a nanoreactor capable of encapsulating gas at pressure up to 1 bar, allowing the investigation of structural evolution under working conditions. To enable atomic-scale observation of nanoparticle structural dynamics, a new type of chip featuring thinner membranes (30 nm) is employed. The nanoreactor is then assembled into the TEM holder and inserted into the TEM chamber. The oxide nanoparticles are initially heated to 400 °C in 10% $O_2$/He to remove contamination. To create a hydrogen oxidation environment, a mixture of 2% $O_2$/20% $H_2$/He (4% $O_2$/20% $H_2$/He, and 10% $O_2$/20% $H_2$/He) is introduced into the nanoreactor, with mass spectrum data collected using the gas released from the nanoreactor. To minimize the influence of water produced during the reaction on structural evolution, we employ a high gas flow rate (0.2 mln/min) to efficiently remove water from the reaction zone.

Sample preparation

$NiFe_2O_4$ nanoparticles (≥99.5%) used for operando TEM experiments were purchased from Alfa Aesar, and the TEM characterization results are shown in **Supplementary Fig. 2 and 3**. $NiFe_2O_4$ has an inverse spinel structure similar to $Fe_3O_4$, with half of the trivalent ions ($Fe^{3+}$) occupying tetrahedral sites, while divalent ions ($Ni^{2+}$, $Fe^{2+}$), and the remaining trivalent iron ions ($Fe^{3+}$) occupy octahedral sites[52]. Upon heating in hydrogen, $NiFe_2O_4$ is reduced to $NiFe-Fe_3O_4$, which can be used to catalyze the hydrogen oxidation reaction.

TEM characterization

TEM images and video series are recorded by a Gatan Rio camera at a 2,048 × 2,048 pixel resolution under an exposure time of 0.05 s. The elemental composition, as well as distribution, are studied on the microscope equipped with an energy-dispersive X-ray analyzer. Electron energy loss spectroscopy (EELS) is collected with Gatan Imaging Filter (GIF) Quantum Model 1077 spectrometer.

Dynamic analysis based on the optical flow field

To quantitatively analyze the LMSI within the in-situ data, we compute the mean displacement vector to measure the extent of the changes between successive frames. The Farnebäck Dense Optical Flow algorithm[48] is utilized to establish the motion field for individual frames. Given our emphasis on the scale of change, the angles of the motion vectors are not considered in this investigation. Consequently, the magnitudes of these motion vectors are leveraged to assess the degree of motion (**Fig. 3i-m**). The magnitudes of the motion vectors of the NiFe catalysts exhibit similar trends of change as the catalytic activity (**Fig. 3a**).

In situ SAED spots tracking

We apply automatic center identification on in-situ SAED data (**Supplementary Video 10**) using the MS-Trans model[53]. The diffraction spots are detected utilizing a Laplacian of Gaussian blob detection algorithm (**Supplementary Video 11**). The elliptical distortion is measured on the first frame using peak finding and ellipse fitting algorithm as mentioned in[49]. The resulting fitted maximum and minimum semi-axes are 422.5 pixels and 421.5 pixels, respectively. The one-pixel discrepancy is likely due to the quantification effects. Therefore, elliptical distortion is neglected in the data processing stage. We monitor the diffraction spots, recording the time each spot appears and disappears to calculate its lifespan. In **Fig. 3n**, we plot the diffraction spots as a function of temperature (x-axis) and reciprocal spacing (y-axis). These spots are indexed to either the $Fe_3O_4$ phase or the NiFe phase, with corresponding crystal planes labeled on the right side of **Fig. 3n**. The spots' different colors and sizes indicate their respective lifespans. As the nanoreactor is raised to reaction temperature, a noticeable decrease in the lifespan of diffraction spots highlights the reaction-related structure dynamics. We select the $Fe_3O_4$ (113) diffraction spots to calculate the average lifespan and assess the structural dynamics of the NiFe-$Fe_3O_4$ catalyst (**Fig. 3o**).

Density functional theory (DFT)

The spin-polarized calculations with the projector-augmented wave (PAW) method are performed based Vienna Ab Initio Simulation package, VASP[54]. The exchange and correlation effects are described by the Perdew–Burke–Ernzerhof functional (PBE)[54,55]. The DFT+U method is used to accurately simulate electron strong correlation in the transition-metal oxide systems[56]. The Hubbard parameters (U) are introduced for the Fe_3d and Ni_3d electrons with the Ueff (= U - J) values of 3.80 eV and 5.77 eV according to previous theoretical work[57,58]. The optimization calculations are employed with a cutoff energy of 400 eV and an atomic force convergence of 0.02 eV/Å. To accurately characterize the magnetic properties of this system, we systematically reference the theoretical studies by Jiao and Wen et al. on $Fe_3O_4$ to appropriately set the spin of octahedral and tetrahedral Fe[59,60]. Among the six possible terminations of the $Fe_3O_4$ (111) surface, the $Fe_{oct2}$-terminated surface is the most stable as hypothesized by STM experiments[61] as well as previous theoretical studies[62]. Therefore, we choose the $Fe_{oct2}$-terminated surface as the model for our studies. For the (111)-$Fe_{oct2}$ model, a p (1 × 1) supercell slab model with 4 atomic-layer thickness is selected. In order to save computer time, 2 atomic layers at the bottom are fixed to simulate the bulk phase, and the top 2 atomic layers are allowed to relax. We apply Monkhorst-Packmesh k points of (3 × 3 × 1) for this surface. For NiFe-$Fe_3O_4$ and Fe-$Fe_3O_4$ models, we use a four layers p (2 × 2) model with a supported NiFe/Fe nanorod, referencing the modeling of interfaces in previous work by Chandler et al[63]. Because of the relatively large cell size, the sampling of the Brillouin zone was restricted to the gamma point. To avoid the interaction between periodic layers, the vacuum layer thickness of both two models is 15 Å.

The free energy for the intermediate is calculated as $\Delta G = \Delta E + \Delta ZPE - T\Delta S$, where $\Delta E$ is the reaction energy change from DFT calculations, $\Delta ZPE$ is the change of zero-point energies (ZPE) which is calculated with the vibrational frequencies of adsorbates and molecules, and $\Delta S$ is the entropy change in the reaction. T is the reaction temperature. Furthermore, all barrier calculations are performed by using the climbing-image nudged elastic band method (NEB) and further improved by the dimer method[64]. The activation energy is calculated by ($E_{TS}$ - $E_{IS}$), where $E_{TS}$ is the energy of the transition state, and $E_{IS}$ is the energy of the initial state.

Molecule dynamic simulation based on neuron network potential

To provide information on LMSI over a larger time and spatial scale, we perform molecule dynamic simulations using a global stochastic surface walking neural network (NN) potential by a commercial software named LASP, (www.lasphub.com) developed by Zhipan Liu group[65]. We construct a NiFe/Fe nanoparticle of 1.5 nm × 1.5 nm (containing 82 atoms), supported by a 5 nm × 4 nm $Fe_3O_4$ (111) substrate (containing 720 Fe atoms and 960 O atoms), to investigate the migration of oxygen in a large-scale model, in total 1762 atoms through theoretical calculations. Molecular dynamic simulation with neural network potential (NN-MD) is performed at a constant temperature (T = 673 K) within the NVT ensemble. We conduct a 20 ps simulation of the NiFe-$Fe_3O_4$ system, which demonstrates that NiFe facilitates the migration of lattice oxygen from the $Fe_3O_4$ to the NiFe nanoparticle. However, in our longer 1000 ps simulation of the Fe system, we observe no migration of lattice oxygen from the $Fe_3O_4$ substrate to the Fe nanoparticle. In our model, the z-axis coordinates of the top layer oxygen atoms of $Fe_3O_4$ are approximately 19.5 Å. Considering the migration effect, we define oxygen atoms with z-axis coordinates greater than 20 Å as oxygen atoms on the nanoparticle, which is consistent with our observations from the MD frames.

**Acknowledgements:**

This work was supported by the National Natural Science Foundation of China (No. U21A20328, 52101277, 22105220, and 22209202), the Strategic Priority Research Program (B) (No. XDB33030200) of Chinese Academy of Sciences, and the Project Funded by China Postdoctoral Science Foundation (No. 2021M703457). We gratefully acknowledge the insightful discussions with Prof. Weixue Li, Zhujun Wang, and Binhang Yan.

**Author contributions:**

Y. P. and D. S. designed the project. Y. P., X. L., and D. Z. performed the operando experiments. J. Z. helped with the sample preparation. Y. P., X. L., M. G., D. Z., and D.